\documentclass{iaus}
\usepackage{bm}
\newcommand{\brac}[1]{\langle #1 \rangle}
\newcommand{\pd}{\partial}

\def\Rs{R_{\odot}}

\usepackage{graphicx}
\usepackage{graphics}
\graphicspath{{fig/}{png/}}

\title[Solar differential rotation] 
{Solar differential rotation: hints to reproduce a near-surface shear 
layer in global simulations}

\author[G. Guerrero, P. Smolarkiewicz, A. Kosovichev \& N. Mansour]  
{G. Guerrero$^1$,  P. K. Smolarkiewicz$^2$,  A. Kosovichev$^1$, 
N. Mansour$^3$} 

\affiliation{$^1$Solar Physics, HEPL, Stanford University, \\ 
452 Lomita mall, Stanford, CA, 94305-4085 \\ 
email: {\tt guerrero@stanford.edu, sasha@sun.stanford.edu} \\[\affilskip]
$^2$National Center for Atmospheric Research, Boulder, CO, 80307, USA \\
 email: {\tt smolar@ucar.edu}\\[\affilskip]
$^3$NASA Ames Research Center, Moffett Field, Mountain View, CA 94040, USA\\
{\tt Nagi.N.Mansour@nasa.gov} 
}

\pubyear{2013}
\volume{294}  
\pagerange{1-8}
\jname{Solar and Astrophysical Dynamos and Magnetic Activity}
\editors{A.G. Kosovichev, E.M. de Gouveia Dal Pino \& Y. Yan}
\begin{document}

\maketitle

\begin{abstract}
Convective turbulent motions in the solar interior, as well as
the mean flows resulting from them, determine the evolution of
the solar magnetic field. With the aim to get a better understanding
of these flows we study anelastic rotating convection 
in a spherical shell whose stratification resembles that of the solar 
interior. This study is done through numerical simulations
performed with the EULAG code. Due to the numerical formulation, 
these simulations are known as implicit large
eddy simulations (ILES), since they intrinsically capture the contribution 
of, non-resolved, small scales at the same time maximizing
the effective Reynolds number.
We reproduce some previous results and find a transition between buoyancy
and rotation dominated regimes which results in anti-solar or solar 
like rotation patterns. Even thought the rotation 
profiles are dominated by Taylor-Proudman columnar rotation, 
we are able to reproduce the tachocline and a low latitude 
near-surface shear layer. We find that simulations results depend 
on the grid resolution as a consequence of a different sub-grid
scale contribution.   
\end{abstract}

\firstsection 
\section{Introduction}

The dynamo mechanism, presumably governing the cyclic evolution of 
the global magnetic field in the Sun (also in late type stars and 
galaxies 
as well as planets), depends on both, the small and large scale 
motions in the solar interior.  Within the 
mean-field theoretical framework, two mechanisms are invoked to 
explain a dynamo cycle. These are so called $\Omega$ and $\alpha$
effects. The first one depends on the differential rotation 
(large scale motion), i.e., the amount of shear able to stretch 
the magnetic field lines. The second effect is the 
contribution of the helical turbulent motions to the amplification 
of magnetic field. It only exists in systems that lack 
reflectional symmetry (i.e., rotating systems). 
Turbulent diffusivity is another important ingredient for a 
dynamo, and other small-scale phenomena like turbulent 
pumping (\cite[Guerrero \& de Gouveia Dal Pino, 2008]{GDP08}) 
or the shear-current 
effect (e.g. \cite[Pipin \& Kosovichev, 2011]{PK11}) 
might also play a role. 

From the large variety of solar dynamo solutions, the Babcock-Leighton 
(BL) flux-transport models are particularly popular since they 
result in magnetic field evolution (sunspot butterfly diagrams) 
that resemble the observations. 
The BL mechanism is a phenomenological formulation of
an $\alpha$-effect that depends on the turbulent diffusion 
of the magnetic flux of active regions and
the poleward transport of magnetic field by meridional flow.
Due to this meridional flow the polar magnetic flux of a previous cycle
is replaced by magnetic flux of opposite polarity. The polar field is then
transported downward towards the base of the convection zone where magnetic
flux tubes are thought to be formed.  However, the 
results of BL and other mean-field models are ambiguous in the sense 
that different combinations of parameters equally resemble the observations. 
Making thus difficult to discern what are the correct ones.
Recent helioseismic results indicate that there may be
two circulation cells per meridional quadrant 
(\cite[see the contribution of Zhao et al. in this proceedings]{ZK12}),
and BL models with such meridional flow profile fail to 
reproduce the observations 
(\cite[Jouve and Brun, 2007]{JB07}). 
Besides, the emergence of magnetic flux tubes from the base of the 
convection zone to the surface is still a matter of debate 
(\cite[Guerrero \& K\"{a}pyl\"{a}, 2011]{GK11}). 

Another attractive alternative, not sufficiently explored so far, 
to explain the solar magnetic cycle is a distributed
dynamo, where the turbulent $\alpha$ effect operates in the entire 
convection zone and not only at the boundaries as in BL models. 
The observed migration pattern corresponds, however, 
to a dynamo wave
traveling equatorwards in a thin layer near the surface where the shear is 
negative (\cite[Brandenburg, 2005]{B05}, \cite[Pipin \& Kosovichev, 2011]{PK11}). 
The rotation of sunspots,
faster than the surface plasma rotation, is one of the arguments in favor of this
alternative.  This model does not depend strongly on the meridional flow
and does not require magnetic field above equipartition as the buoyancy 
of flux tubes model does. Explaining the sunspot formation and their properties 
in this model is still problematic, but these phenomena are in general
still an open question and deserve further study.

To advance our understanding of the solar dynamo mechanism it is important
to understand first the dynamics of the flows in the convection zone.  From a
theoretical point of view, the most realistic approach 
to study the solar flows is through global numerical simulations (GNS).
Although the current computing capabilities are unable to resolve the
essential scales in the Sun,
GNS are able to study systems that resemble some of the solar properties. The
main expectation is to get closer to the reality and understand the complicated
multi-scale physics. Since the first 
global numerical models (e.g. \cite[Gilman, 1977]{G77}) there has been huge
progress in the understanding of convection in rotating spherical shells, 
but till now, {\it ab-initio} 3D numerical models still have 
difficulties in reproducing 
the observed properties of the solar differential rotation 
(e.g. \cite[Miesch et al., 2006]{MBT06}, 
\cite[K\"{a}pyl\"{a} et al. 2011]{KMGBC11}). 
To our knowledge just a few groups are addressing this problem.  The ASH 
group (e.g., \cite[Miesch et al., 2006]{MBT06}) uses an anelastic, 
spectral code (as well as \cite[Busse \& Simitiev, 2010]{BS11}), 
the Pencil-Code group (e.g., \cite[K\"{a}pyl\"{a} et al. 2011]{KMGBC11}) 
which  uses finite differences to perform compressible simulations in 
a wedge geometry and the EULAG 
group (e.g., \cite[Ghizaru et al., 2010]{GC10}), which so far 
has focused on the solar dynamo rather than on 
the differential rotation problem.

In this paper we present the initial results of our attempt to address
the differential rotation problem with the EULAG code 
(\cite[Smolarkiewicz et al., 2001]{SMW01}, \cite[Prusa et al., 2008]{PEW08}). 
EULAG is an anelastic code for geosphysical and astrophysical fluid 
dynamics, with unique semi-implicit numerics built on high-resolution 
nonoscillatory forward-in-time (NFT) advection schemes MPDATA 
(for {\it multidimensional positive definite advection transport 
algorithm}); cf. \cite[Smolarkiewicz, 2006]{S06} for a recent 
overview.
The code does not require any explicit viscosity in order to
remain stable, and the numerical viscosity has been identified to 
work similarly to the eddy viscosity used in different sub-grid 
scale models (\cite[Margolin \& Rider, 2002]{MR02}). 
For this reason EULAG results are interpreted as
ILES (\cite[Smolarkiewicz \& Margolin, 2007]{SM07}). 

\section{Model setup}

We solve the anelastic set of hydrodynamic equations following
the formulation of \cite{LH82} and \cite{L90}:
\begin{equation}
          {\bm \nabla}\cdot(\rho_s\bm u)=0, \label{equ:cont}
\end{equation}
\begin{equation}
  \frac{D \bm u}{Dt}= 2{\bm \Omega} \times {\bm u}   
  + {\bf g}\frac{\Theta'}
  {\Theta_s} -{\bm \nabla}\left(\frac{p'}{\rho_s}\right) \;, \label{equ:mom} 
\end{equation}
\begin{equation}
 \frac{D \Theta'}{Dt} = -{\bm u}\cdot {\bm \nabla}\Theta_e + \frac{1}{\rho_s}
                        {\cal H}(\Theta')-\alpha\Theta'
 \label{eq:en} 
\end{equation}

where $D/Dt = \pd/\pd t + \bm{u} \cdot {\bm \nabla}$ is the total
time derivative, ${\bm u}$, is the velocity,  $p'$ and $\Theta'$ are the
pressure and potential temperature fluctuations, respectively and
$\cal{H}(\theta')$ absorbs the radiative and heat diffusion terms.
$\rho_s$  and $\Theta_s$ are the density and potential temperature 
of the reference state chosen to be isentropic.  
The radial profile of $\rho_s$ is defined considering 
hydrostatic equilibrium with $g\propto1/r^2$ and 
$\Theta_s={\rm const}$. The potential temperature is related to the
specific entropy via $s=c_p \ln\Theta+{\rm const}$. 
In Eq. ~(\ref{equ:mom}) ${\bm \Omega}$ is the rotation vector.
Finally, $\Theta_e$ is
the potential temperature of an ambient state which is assumed to 
be an azimuthal and long-term temporal average of a static solution
of the system.  The term $\alpha\Theta'$ in eq. 
(\ref{eq:en}), forces
the system towards the ambient state, $\Theta_e$, on a timescale
$\tau=\alpha^{-1}=1.55{\rm e}8$ s. This timescale is much shorter than
the timescale of radiative and heat diffusion, meaning that this
term ultimately drives convection in the system.
The fact that the Sun does not exhibit
important changes in its thermal state over the time scale of our 
interest,  make us to believe that a 1D solar structural model is 
a good  approximation for the ambient state of the solar interior. 

For the simulations presented here we approximate the ambient state 
by a polytropic model. Our domain spans in radius from 
$0.62 \Rs$ to $0.96\Rs$ (although in the final section we 
extend our domain up to $r=0.985\Rs$). From the bottom up to $r=0.71\Rs$ 
the model is stable to convection with polytropic index
$m=3$. The upper layer is convectively unstable with $m=1.49995$. 
In latitude and longitude we consider $0\le \theta \le \pi$ and 
$0 \le \phi \le 2\pi$, respectively. 
In the latitudinal direction discrete differentiation extends 
across the poles, while flipping sign of the latitudinal and 
meridional components of differentiated vector fields.
In the radial direction we use stress
free, impermeable, boundary conditions for the velocity field, 
whereas for $\Theta'$ zero normal derivative is assumed.

\section{Results}

We have performed spherical shell convection simulations aiming to
reproduce the solar differential rotation. As mentioned in the previous
section, our numerical model does not explicitly include dissipative terms. 
For this reason, it is not easy to determine precise values of non-dimensional
quantities like the Reynolds, Rayleight or Taylor numbers
(\cite[Domaradzki et al., 2003]{DXS03}).
For this paper we characterize our simulations with the rotation rate and the
grid resolution. 

\subsection{Convection vs. rotation}

\begin{figure}[htb]
\begin{center}
\includegraphics[width=0.48\columnwidth]{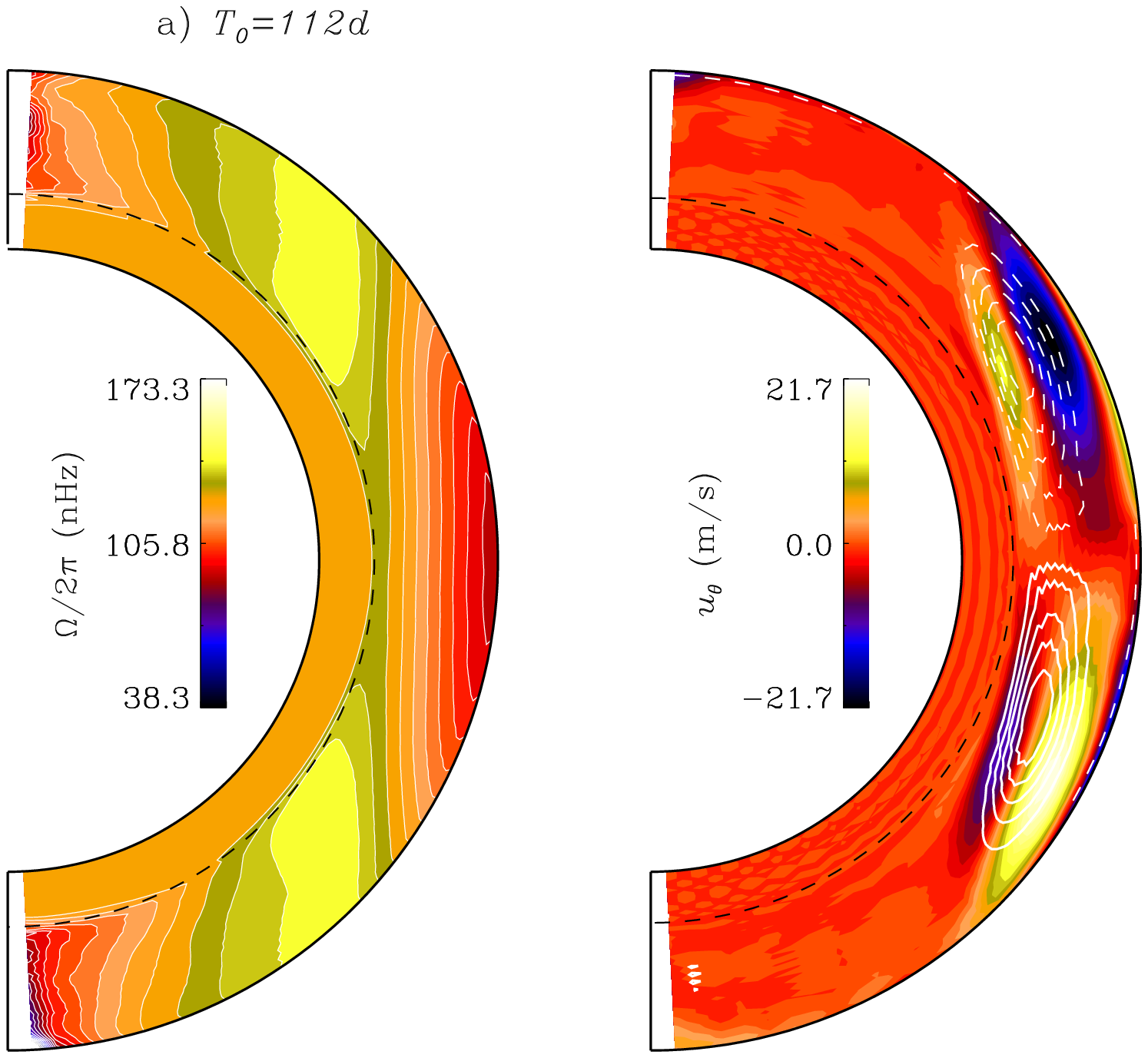}\hspace{0.4 cm}
\includegraphics[width=0.48\columnwidth]{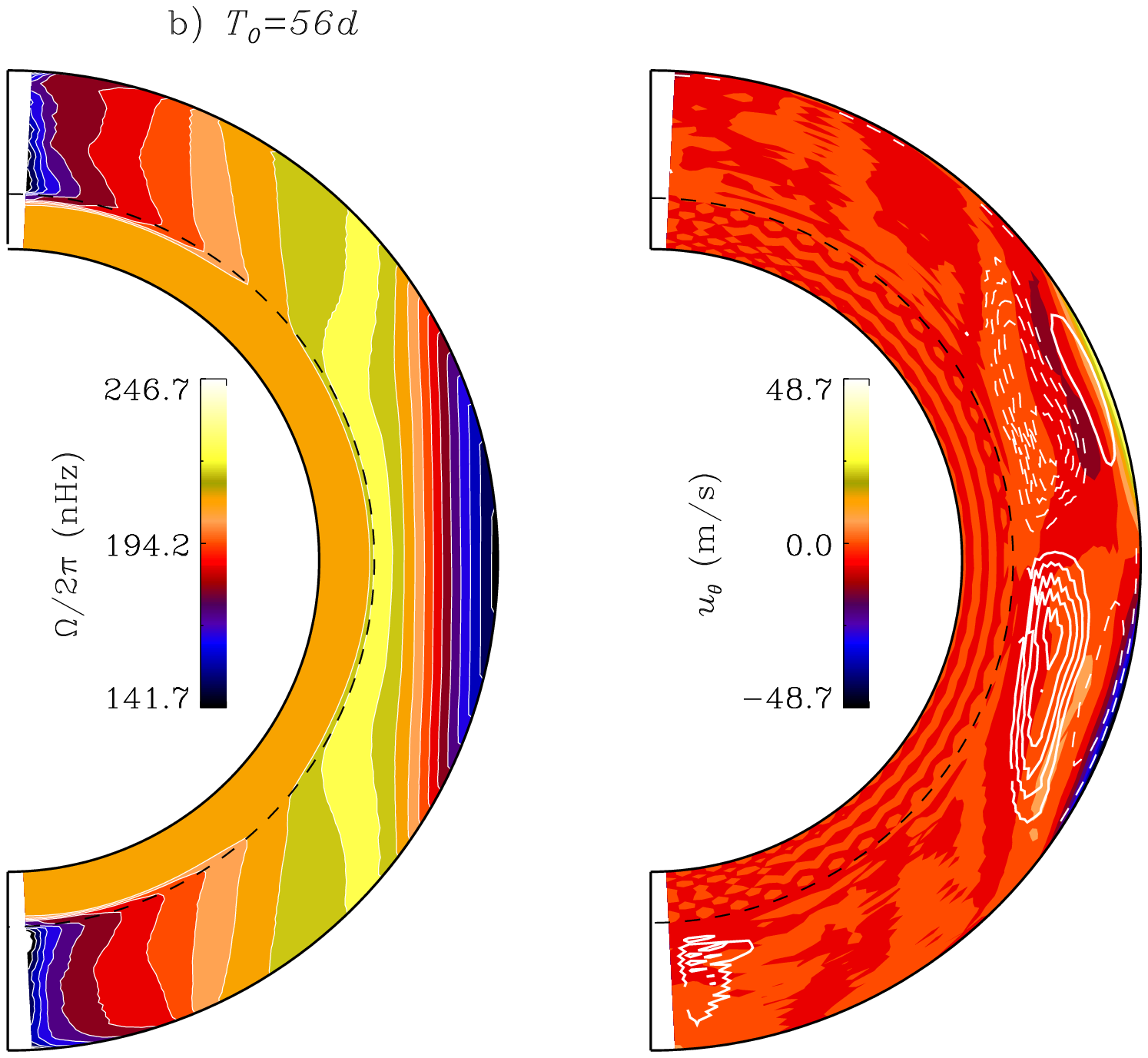}\\
\includegraphics[width=0.48\columnwidth]{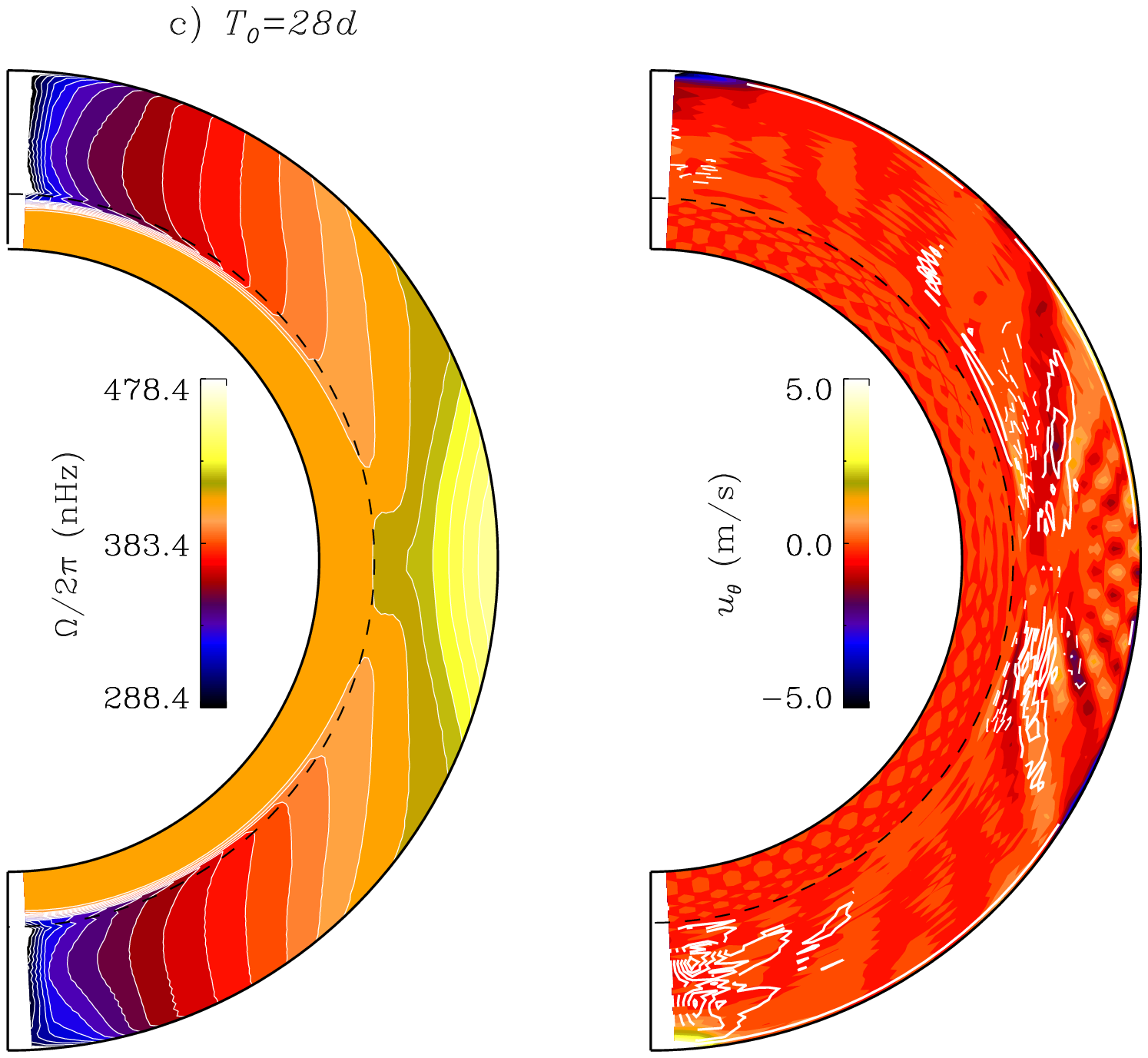}\hspace{0.4 cm}
\includegraphics[width=0.48\columnwidth]{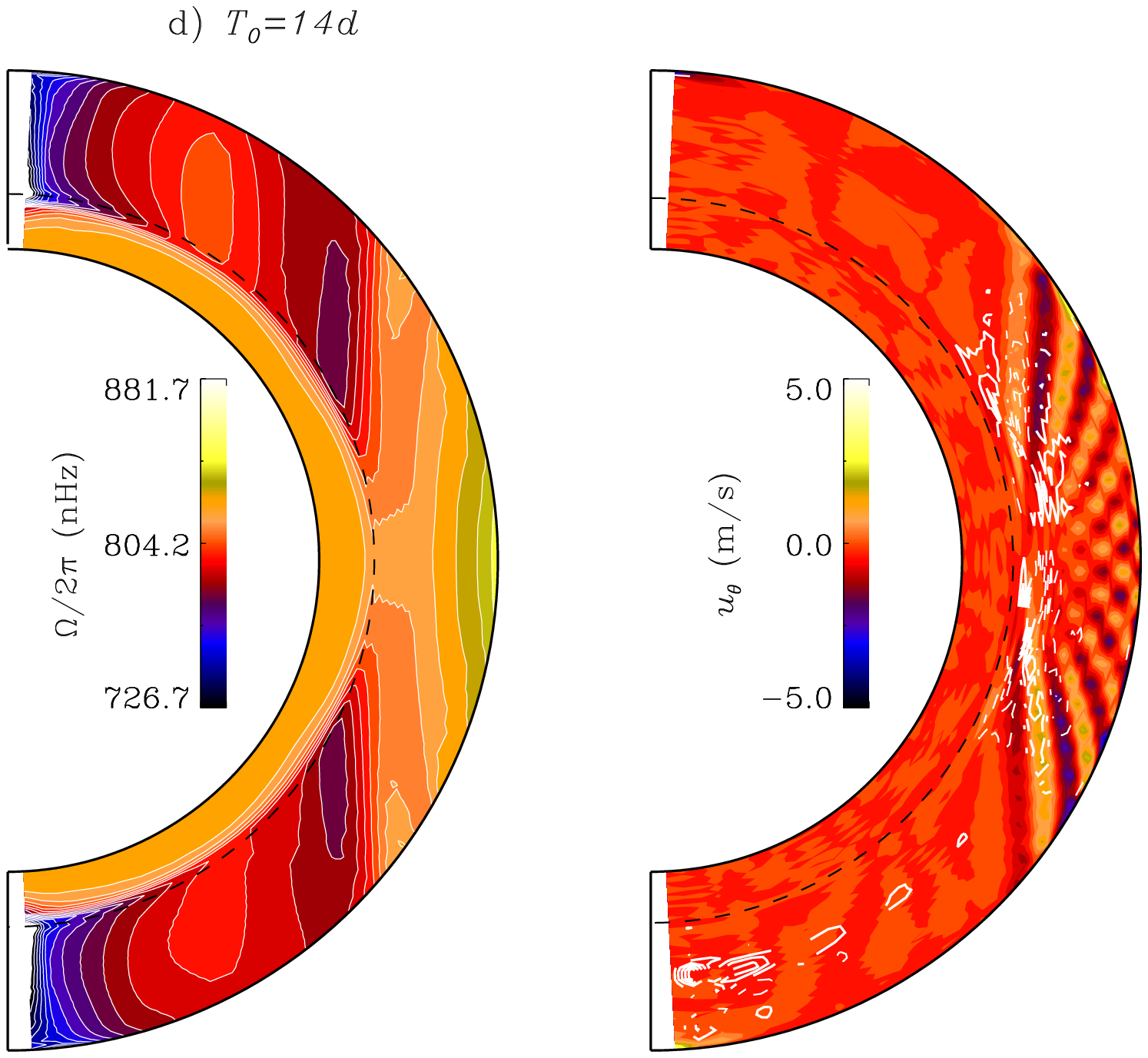}
\caption{Differential rotation and meridional circulation profiles 
for the models with
different rotation period, $T_0$, as indicated in the legend. The profiles
correspond to longitude-time averages computed in the, statistically, 
steady stage during a 3-years time span. Continuous (dashed)
lines depict clockwise (counterclockwise) contours of the stream flow.
They represent the meridional circulation profile.}
\label{fig1}
\end{center}
\end{figure}

In the first set of simulations we consider models with
different rotation period and keep constant all other parameters. 
We study cases with $T_0=112d$ ($\Omega_0=0.25 \Omega_{\odot}$,  
(case L0.25), $T_0=56d$ ($0.5\Omega_{\odot}$, L0.5),
$T_0=28d$ ($\Omega_{\odot}$, L1) and $T_0=14d$ ($2\Omega_{\odot}$, L2).
These simulations are performed with a rather coarse resolution 
($n_{\phi}=128$, $n_{\theta}=64$, $n_r=47$) previously adopted by
\cite[Ghizaru et al. (2010)]{GC10}.
The results are shown in Fig. \ref{fig1}.

As it was described in 
\cite{G77}, the resulting rotation pattern is defined by the 
ratio between the rotation rate and the amplitude of the 
convective velocities, i.e., by the competition between 
Coriolis vs. buoyant forces. 
Our results agree with previous studies of rotating convection 
(\cite[K\"{a}pyl\"{a} et al. 2011]{KMGBC11}).
For the slow rotating cases buoyancy dominates,
the correlation between the azimuthal and the vertical velocities,
$R_{r \phi}=\brac{u_{r}' u_{\phi}'}$,
is mainly negative so that there is an inwards transfer of angular
momentum. Probably due to the meridional Reynolds stress component,
$R_{r \theta}=\brac{u_{r}' u_{\theta}'}$,
a single meridional circulation cell at each meridional 
quadrant is formed. It is counterclockwise 
in the northern and clockwise in the southern hemispheres. 
This circulation, in turn, redistributes via advection the angular momentum
leading to faster rotation at higher latitudes.  

With the increase of the
rotation rate, Coriolis forces dominates over buoyancy.  The latitudinal 
component of the Reynolds stress 
tensor, $R_{\theta \phi}=\brac{u_{\theta}' u_{\phi}'}$, results positive 
(negative) in the northern (southern) hemisphere and tends to 
concentrate at lower 
latitudes at the base of the convection zone (\cite[Guerrero et al. (2013), 
in preparation]{GSK13}).  The tensor component 
$R_{r\phi}$, which is symmetric across the equator,
changes sign for $\Omega_0\ge\Omega_{\odot}$, switching the rotation
profile from decelerated to accelerated in the equatorial region.
In these cases the meridional circulation is multicellular and
does not seem to significantly affect the distribution of 
angular momentum.
The contrast in rotation between the equator and $60^{\circ}$ latitude 
in case L1 is $\Delta\Omega\simeq80$ nHz, in good agreement with the Sun 
where this difference is $\sim 85$ nHz. 

As it has been found in previous solar simulations with EULAG 
(i.e. \cite[Ghizaru et al. 2010]{GCS10}, \cite[Racine et al. 2011]
{RCGBS11}), the rotation profile exhibits  a 
well defined tachocline. It appears due to a strongly sub-adiabatic
layer ($r<0.71\Rs$) obtained with a steeper profile of $\Theta_e$.  
Numerical experiments not included here,
with a less steeper $\Theta_e$ profile between the two regions,
result in the transfer of angular momentum towards the stable region.
The mean flows resulting from those models dramatically differ
from the ones reported here. 

The profiles of differential rotation for most of the simulations
show alignment of iso-rotation lines along 
the rotation axis (Taylor-Proudman theorem). For the
case L1, a slight departure from this columnar distribution starts
to appear at latitudes above $\sim70^{\circ}$. A solar-like profile,
with conical contours of iso-rotation could be achieved with the
fine setting of some model parameters (Charbonneau et al. 2012, 
private communication).  

The meridional flow, depicted in Fig.  \ref{fig1} with continuous 
lines, corresponds to a clockwise circulation, dashed lines correspond
to counterclockwise circulation. It varies from a single,
coherent, cell at each hemisphere for the case L0.25 to a double
cell configuration for case L0.5 and a multicellular, not well 
defined, pattern for cases L1 and L2 ($\Omega_0 \ge \Omega_{\odot}$).   

\subsection{Convergence experiments}
In the next set of numerical experiments we 
keep the period of rotation fixed to the solar value
($\Omega_0=\Omega_{\odot}$) and explore the effects of the implicit
LES viscosity and the convergence of the 
results by increasing the resolution. We denote these 
models as case L1 ($n_{\phi}=128$, $n_{\theta}=64$, $n_r=47$), M1
($n_{\phi}=256$, $n_{\theta}=128$, $n_r=94$)
and H1 ($n_{\phi}=512$, $n_{\theta}=256$, $n_r=188$). 

\begin{figure}[!ht]
\begin{center}
\includegraphics[width=0.48\columnwidth]{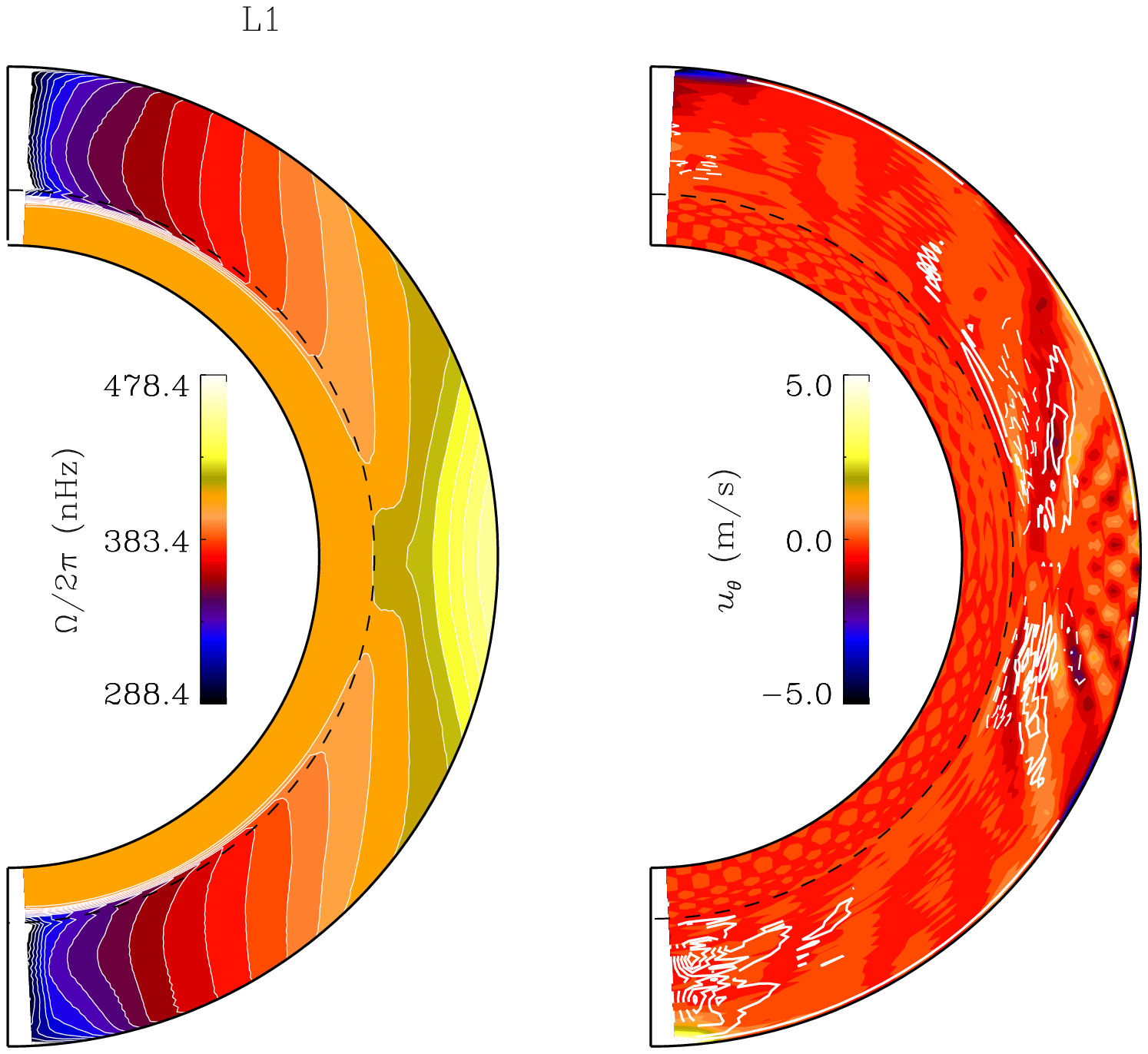}\hspace{0.1 cm}
\includegraphics[width=0.48\columnwidth]{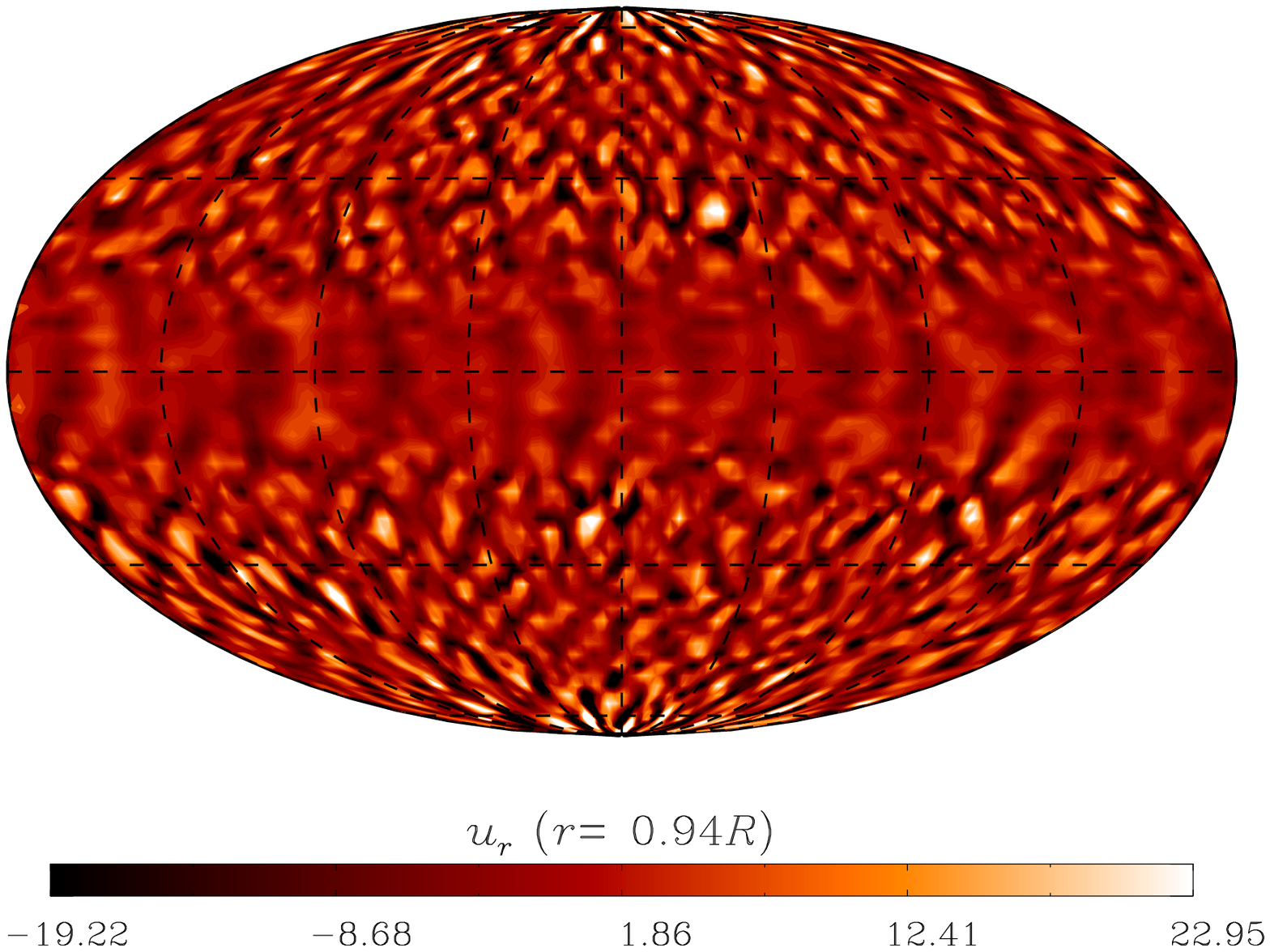}\\
\includegraphics[width=0.48\columnwidth]{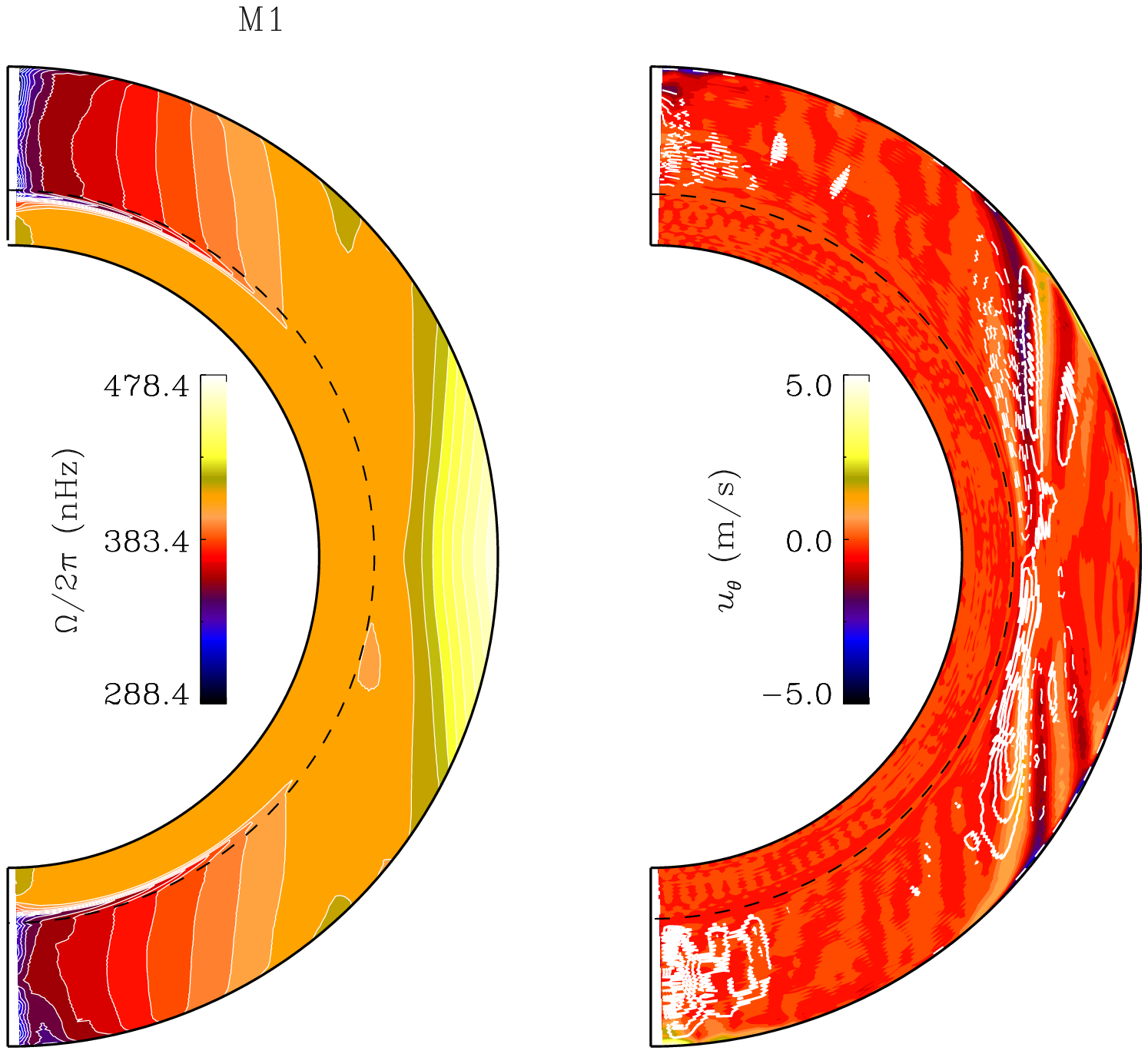}\hspace{0.1 cm}
\includegraphics[width=0.48\columnwidth]{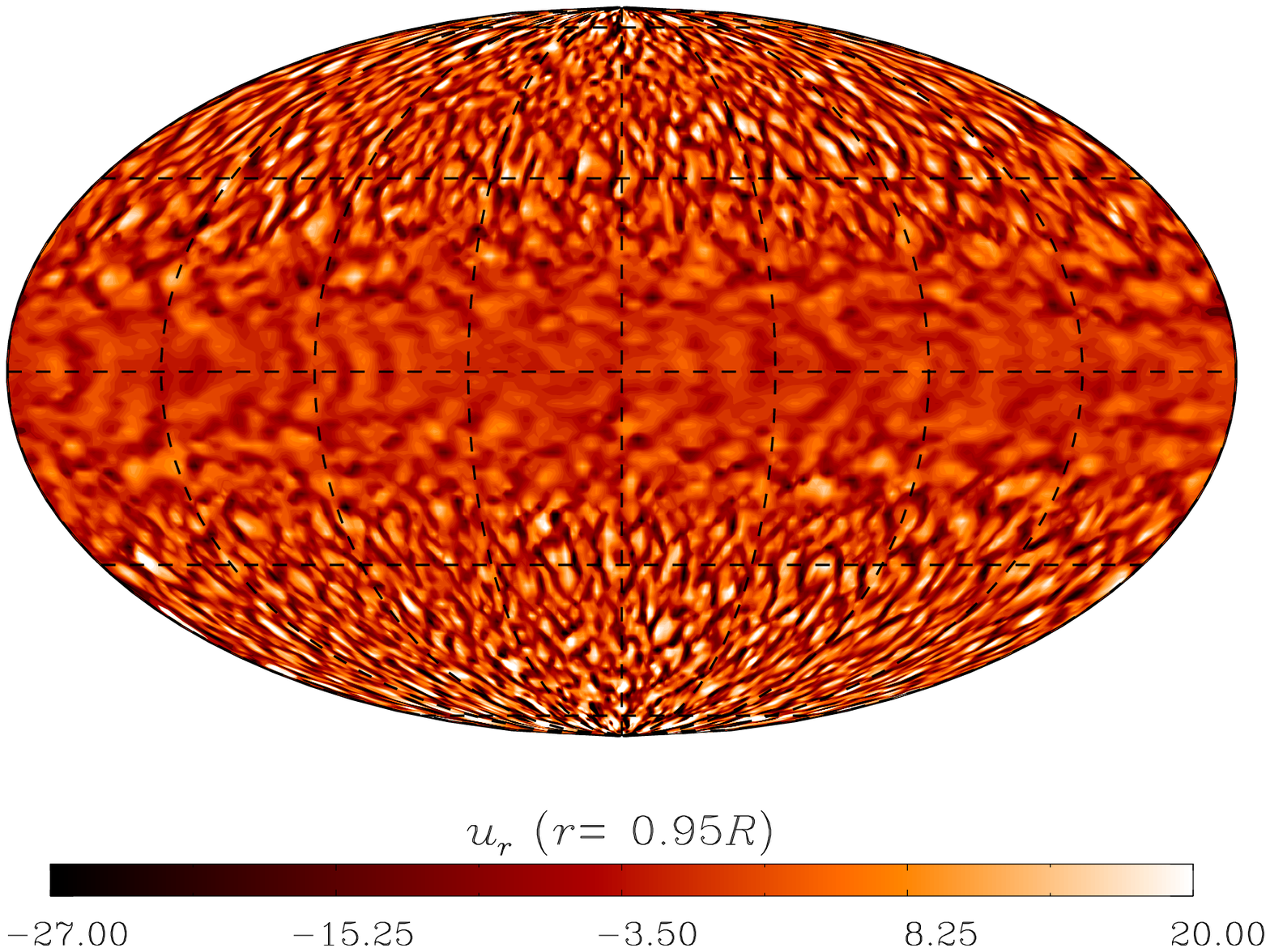}\\
\includegraphics[width=0.48\columnwidth]{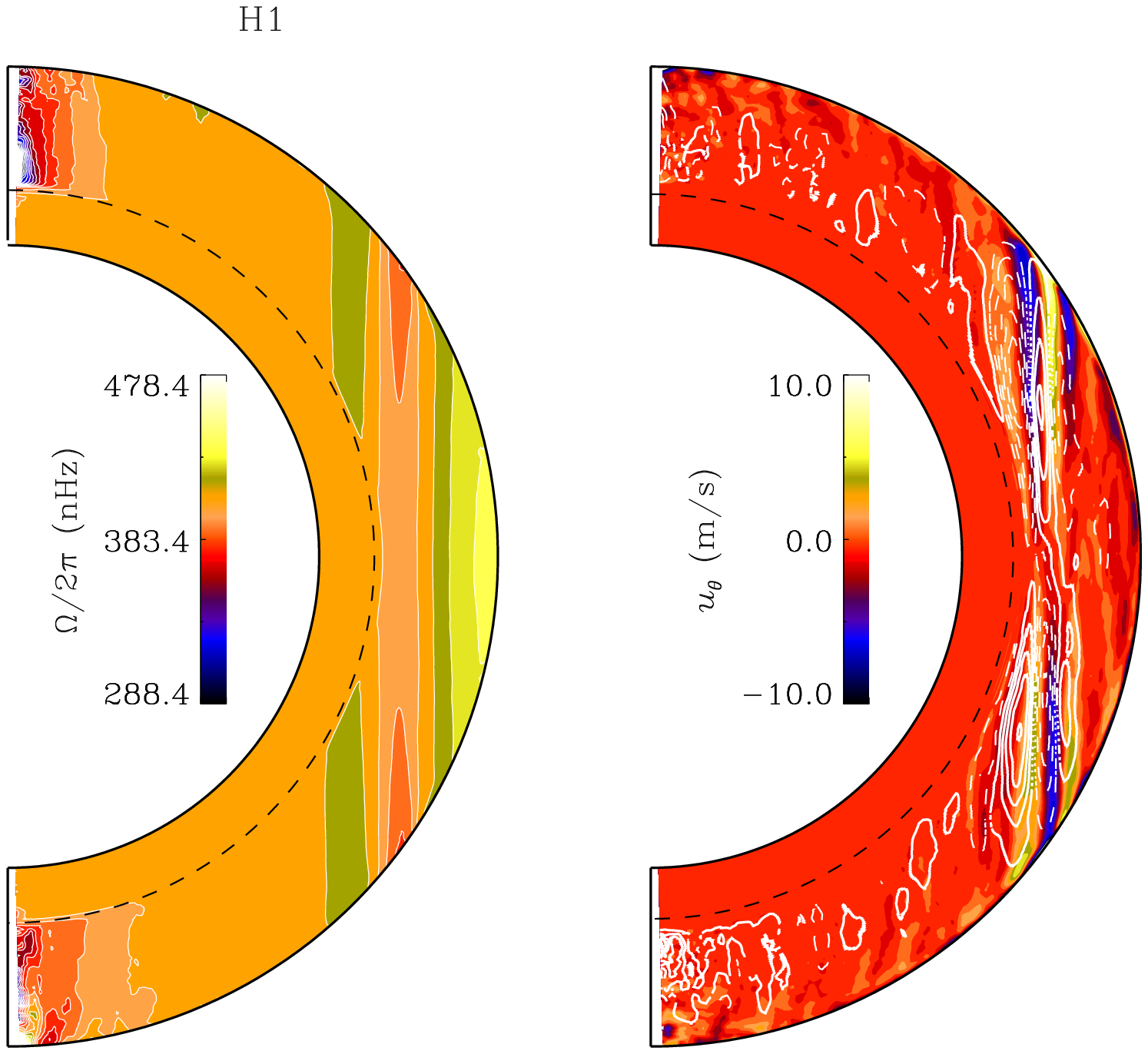}\hspace{0.1 cm}
\includegraphics[width=0.48\columnwidth]{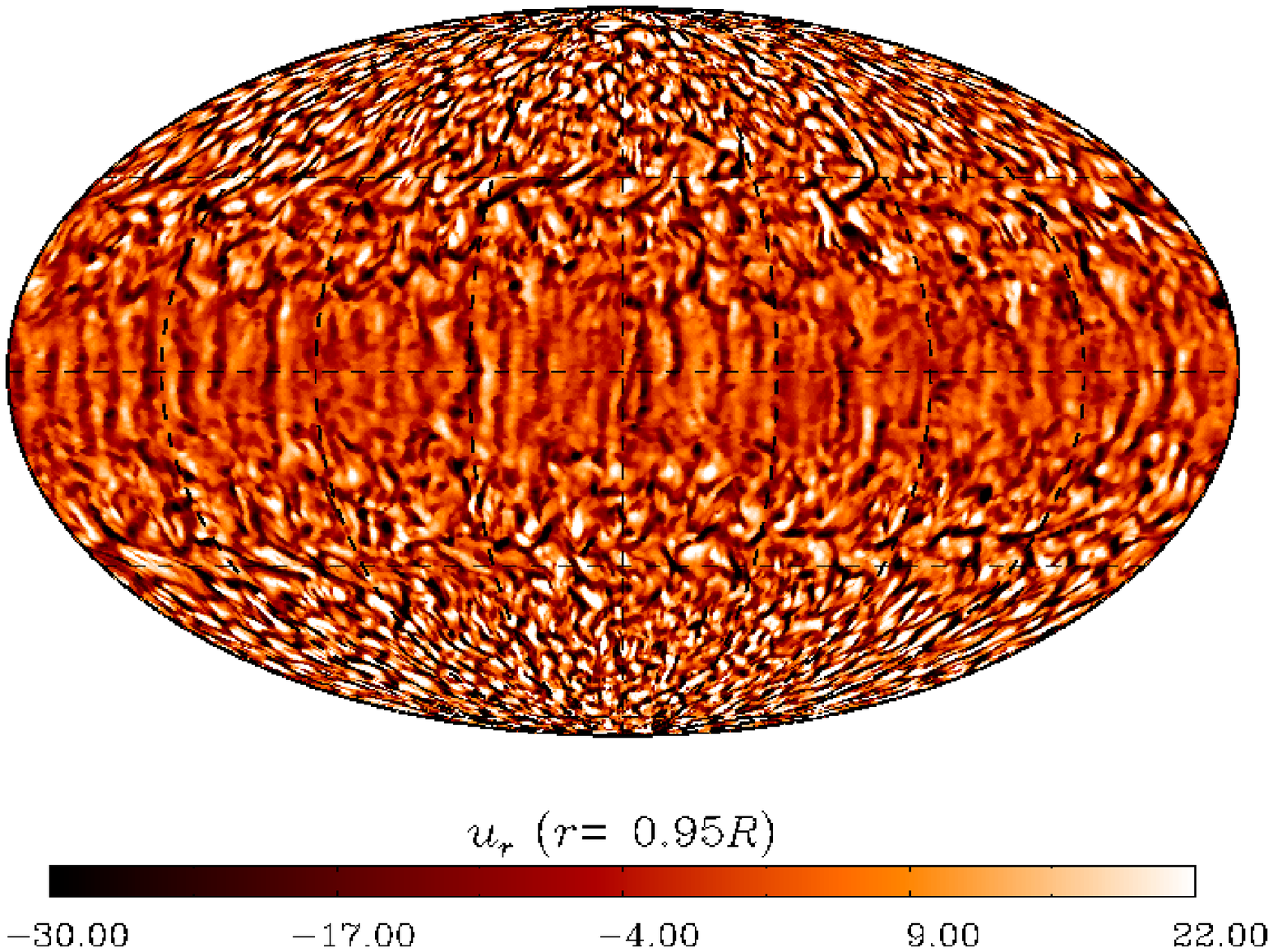}\\
\caption{Left: differential rotation profiles for models with different 
numerical resolution. Models L1  ($n_{\phi}=128$, $n_{\theta}=64$, $n_r=47$),
M1 ($n_{\phi}=256$, $n_{\theta}=128$, $n_r=94$), and H1 
($n_{\phi}=512$, $n_{\theta}=256$, $n_r=188$)  are shown from top 
to bottom. Middle: meridional circulation profiles. 
Right: vertical velocity (in m/s) at the top of the domain.}
\label{fig2} 
\end{center}
\end{figure}

For all the cases the results show a solar like rotation, i.e., faster
in the equatorial regions and slower rotation at the poles.
However, only in the lower resolution case
there is a monotonic decrease of the rotation rates from the equator
to the poles.  The model ~M1 shows a faster equator, an extended region
of iso-rotation with the stable layer and slower poles. The contrast
between the equator and $60^{\circ}$ is $\sim 70$nHz.
The model ~H1 shows an additional  
column of slower rotation at intermediate latitudes followed by
an extended region of iso-rotation. In this case 
$\Delta \Omega\simeq 20$ nHz. All models show a multicellular 
pattern of meridional flow. 
The right hand side panels of Fig. \ref{fig2} show the distribution
of radial velocities at the top of the domain for models with 
different grid resolution.  Although there is a clear difference
between the scales that each model is able to resolve, 
the presence of "banana cells" in a belt of $\pm 30^{\circ}$
around the equator is common in all simulations. At higher latitudes 
the convection cells have small spatial extent. 

The differences between the three cases are possibly 
due to insufficient relaxation time, especially for the case H1. 
Furthermore, with the increased resolution new modes of 
motion appear and the role of subgrid-scale transport diminishes. 
Consequently, reproducing features of the low resolution result 
at higher resolution, may require incorporating explicit eddy 
transport or readjustment of the parameterized
turbulent heat flux (i.e., $\Theta_e$ and the rate of
Newtonian cooling). We remark here 
that the model is sensitive to changes in these parameters. 
Further investigation of solution sensitivities to explicit 
and implicit SGS viscosities and their interplay with means 
of forcing convection will be important to determine an 
optimal way of modeling small scale contributions to the
global problem.

\subsection{Near-surface shear layer}

From the previous simulations we noticed that when convection is 
vigorous enough, such that its time-scale is shorter than the 
rotation period, the Coriolis force ceases to effectively deflect
the radial upflows and downflows.
This results in negative values of 
the Reynold stress component $\brac{u_r' u_{\phi}'}$ and, consequently, in 
an inwards flux of angular momentum (i.e., decrease in the rotation 
rate). This is believed to occur in the upper layers of the 
solar convection zone,
where the time scales of granulation (minutes) and supergranulation 
(8 - 24 hours) are much shorter than the rotation period (28 days). 
This physical mechanism is thought to be, at least in part, 
responsible for the formation of the near-surface shear layer 
(\cite[Miesch \& Hindman, 2011]{MH11}) and has been verified in
global simulations of the uppermost fraction of the convection zone
(\cite[DeRosa et al. 2002]{DGT02}).   
\begin{figure}[htb]
\begin{center}
\includegraphics[width=0.48\columnwidth]{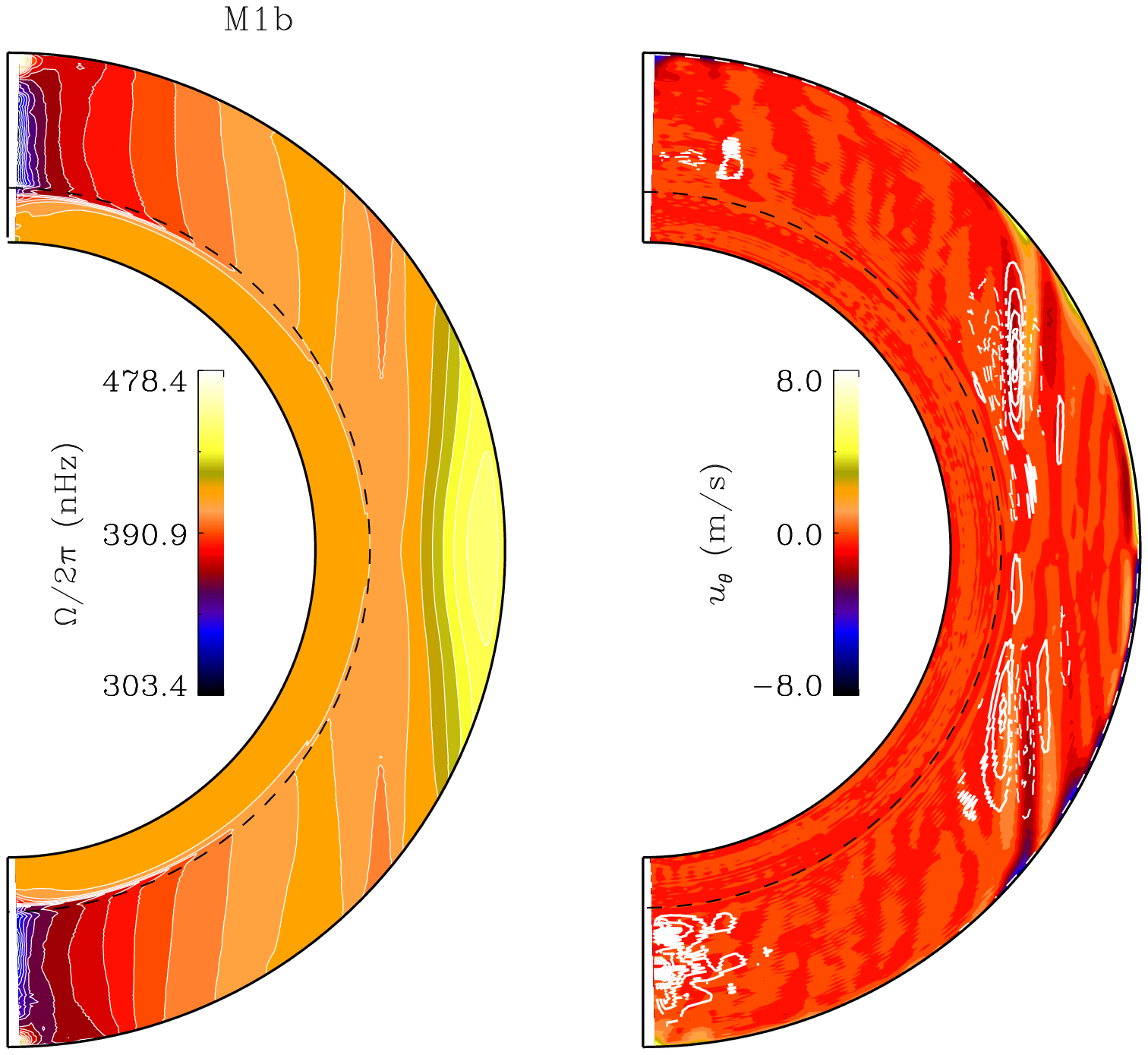}\hspace{0.1 cm}
\includegraphics[width=0.48\columnwidth]{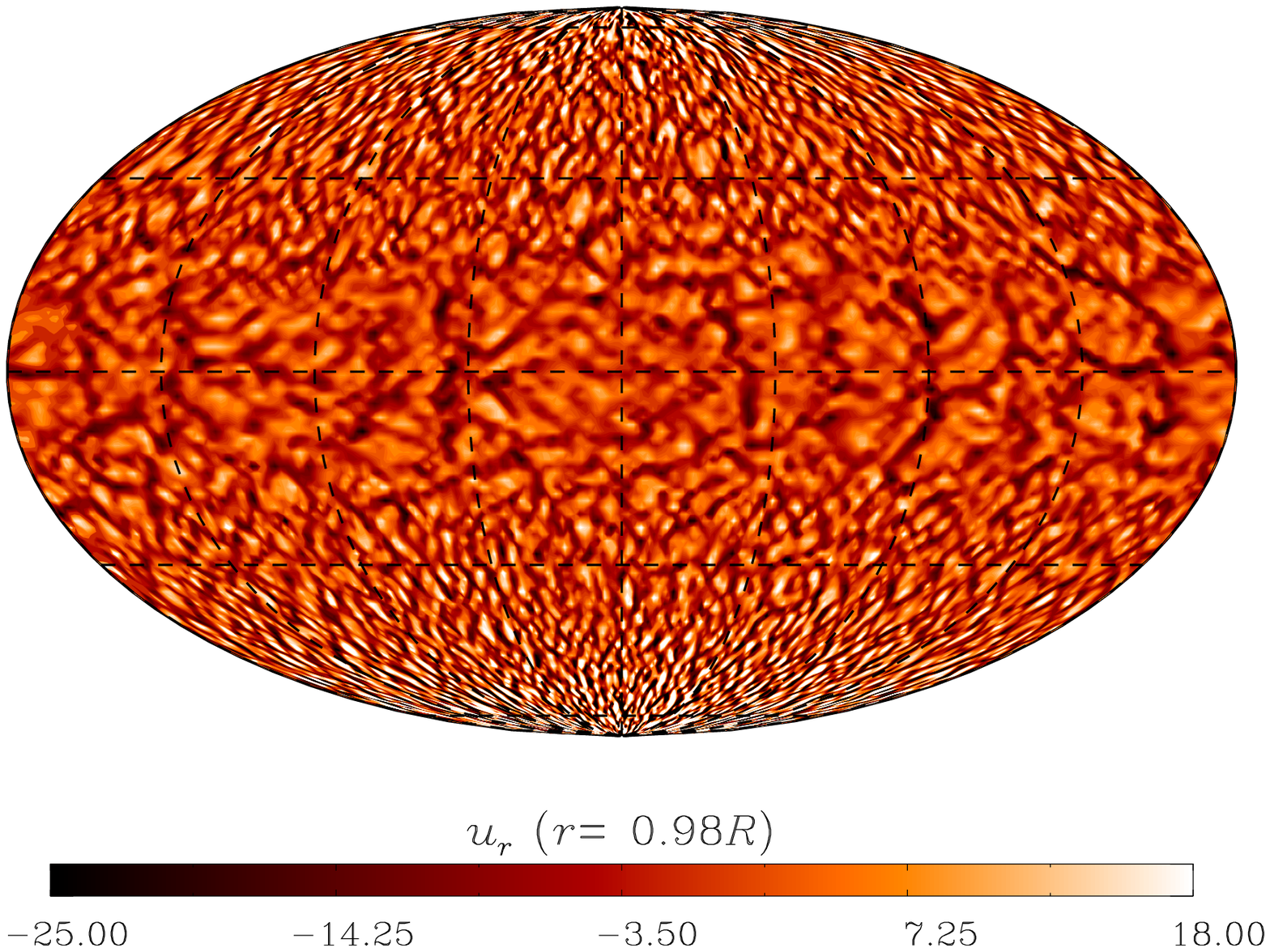}\\
\caption{Same than Fig. \ref{fig2} for the model M1b. In this case
the vertical domain extends up to $r=0.985\Rs$.}
\label{fig3}
\end{center}
\end{figure}

In order to include two different convection regimes (one dominated
by rotation and another dominated by buoyancy) in the 
convection zone of our model, we extend the domain in the 
radial direction up to $r=0.985\Rs$ and
add a third layer with the polytropic index slightly smaller than the
one in the convection zone, $m_1=1.4999$. Thus, the ambient state 
$\Theta_e$ decreases
faster (increasing superadiabaticity) in the upper 5\% of the domain 
(compare dotted line in Fig. \ref{fig3}a and b). 

In Fig. \ref{fig3} we present the differential rotation and meridional
circulation profiles for this model (case ~M1b).  The rotation profile
is similar to the case ~M1, however in this case 
$\Delta\Theta$ is only  $\simeq 30$ nHz. The meridional flow exhibit
multiple cells.  The radial velocity at the surface level shows
broad convective structures at lower latitudes and small convection
cells at middle and higher latitudes.  The equatorial structures
are still elongated in latitude, however the banana cells are not 
evident at this height. Like in the run ~M1, they are evident
 at $r=0.96\Rs$.

\begin{figure}[htb]
\begin{center}
\includegraphics[width=0.49\columnwidth]{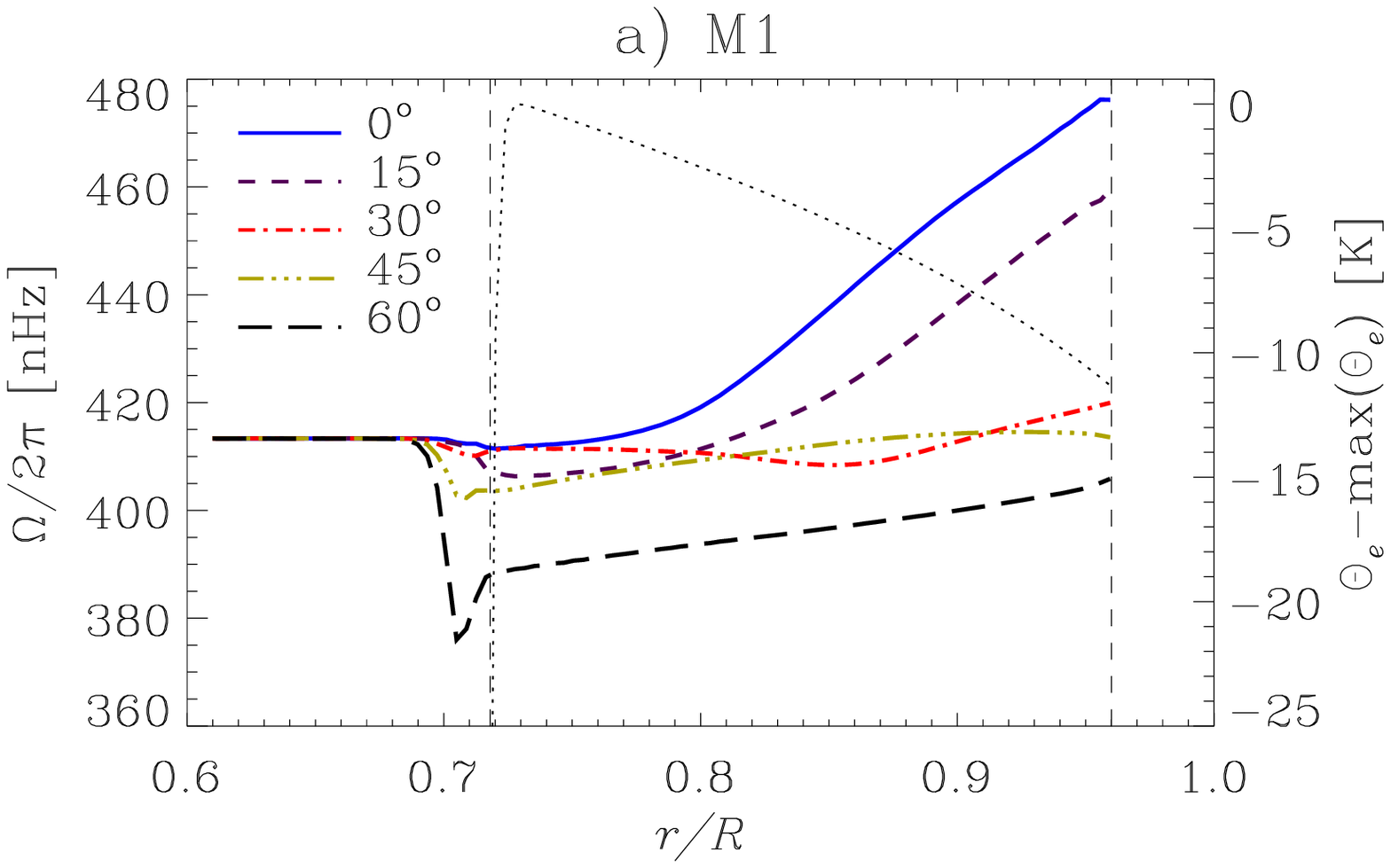}\hspace{0.05 cm}
\includegraphics[width=0.49\columnwidth]{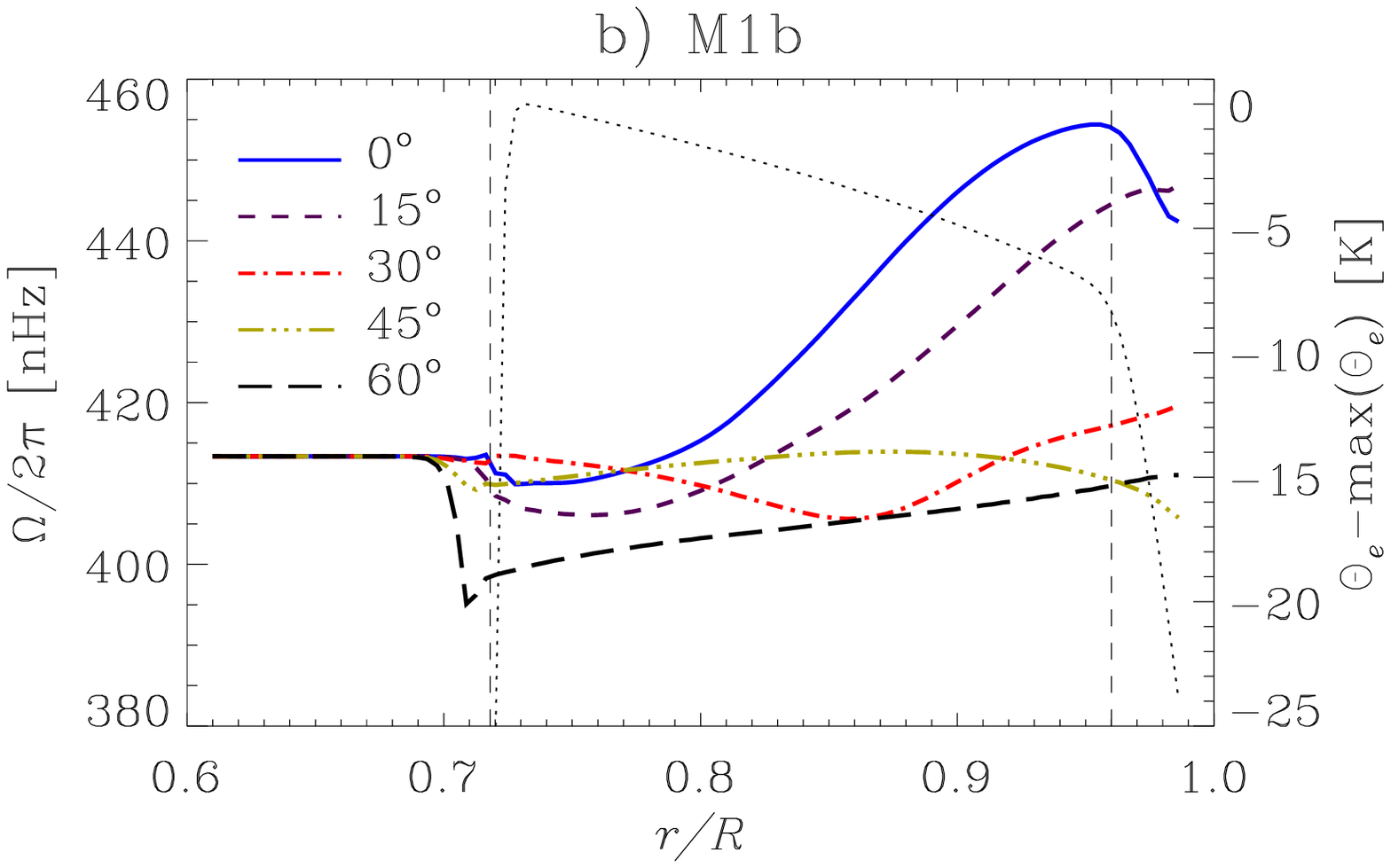}
\caption{Radial distribution of the angular velocity for different 
co-latitudes. Left and right hand panels correspond to the models 
M1 and the M1b, respectively. A near surface shear layer appears 
for $r>0.96R$. 
The thin-dotted line corresponds to the radial
distribution of $\Theta_e$ scaled with its maximum value.}
\label{fig4}
\end{center}
\end{figure}

In Fig. \ref{fig4} we compare the radial distribution of 
differential rotation for the models
without (M1) and with (M1b) this additional layer.
Although the radial profiles of differential rotation 
for the case ~M1b slightly differ from that without 
the upper layer (~M1),  negative 
shear at lower latitudes is clearly observed (similar results were 
found by \cite[K\"{a}pyl\"{a} et al., 2011]{KMB11} in simulations 
with a large density contrast between the bottom and top of the
domain).  We notice that the radial component
of the Reynolds stress tensor is also negative at 
these latitudes. However, our simulations are unable to
capture the dynamics of the surface region and could not
be extended above $r=0.985\Rs$. An extended region of negative
angular momentum flux might result in a pronounced poleward meridional
flow near the surface, not observed here but present in the Sun. 
This meridional flow could transport angular momentum to higher 
latitudes and thus form a near-surface shear layer similar to the
observations, as suggested by \cite{MH11}. 

\section{Conclusions}

We have used the anelastic, hydrodynamic, version of the EULAG 
code to perform global numerical
simulations of convection in a rotating stratified envelope and study
different regimes of differential rotation.
We consider first models with
a bottom stable (subadiabatic) region and an extended 
convectively unstable layer. In these models we are able
to reproduce previous results obtained with different
codes. Different correlations between the turbulent 
velocities (Reynold stresses) are found for models with 
different rotation rates. These correlations appear
as the competition between buoyancy and Coriolis
forces and give rise to different mean flows patterns.
Models in which convective velocities dominate over the rotation
velocity result in profiles of the angular velocity with the equator
rotating slower than higher latitudes. Models in which rotation
dominates result in a faster equator and slower poles. 
For all models the bottom, convectively stable, layer rotates 
uniformly forming a region of strong rotational shear (tachocline) 
at the base  of the convection zone. Unlike the Sun in all the models 
the rotation contours are mainly aligned along the rotation axis 
(cylindrical) following the Taylor-Proudman theorem.

We have also run simulations for models with higher numerical 
resolution, and obtained the mean flow profiles somewhat
different from the lower resolution base model (L1). 
Paradoxically, the model with the coarse grid resembles 
better the solar rotation profile than the models with 
finer grids.  This emphasizes the importance of the 
balance between the efficiency of convective mixing at 
resolved scales, the implicit eddy viscosities and 
the large scale forcing. 


Finally, to model the fast convective motions at the top of the
convection zone, we have added a third layer in the top of 
the domain where the potential temperature (entropy) declines
quickly (corresponding to an increase of super adiabaticity).  
This layer generates convective motions with
smaller spatial and temporal scales less affected by
rotation than the deeper slow motions. In this model a near 
surface shear layer is
formed at lower latitudes due to the negative radial
transfer of the angular momentum. \\

\noindent {\bf Acknowledgements}
We thank P. Charbonneau and J-F Cossette for their important
help in the construction of the spherical model.
GG acknowledges NSF for travel support. All the simulation
here were performed in the NASA cluster Pleiades.

\end{document}